# Primary analysis method for incomplete CD4 count data from IMPI trial and other trials with similar setting.

Abdul-Karim Iddrisu[1], Abukari Alhassan[2]

[1]University of Energy and Natural Resources, Department of Mathematics and Statistics, Ghana.
[2]University for Development Studies, Department of Statistics, Ghana.

ARTICLE INFO

Received 00 January  00
Accepted 00 January 00

Correspondence to:
Abdul-Karim Iddrisu, PhD, MSc, PgDs, and BSc.,
Department of Mathematics and Statistics, University of Energy and Natural Resources, Box 214, Suntan, Bono Region, Ghana
E-mail: abdul-karim.iddrisu@uenr.edu.gh

ORCID IDs:
Abdul-Karim Iddrisu:
https://orcid.org/0000-0002-6751-2516
Abukari Alhassan:
https://orcid.org/0000-0001-7845-9768

Fuding:
This study receives no funding.
Conflict of Interest
No potential conflict of interest relevant to this article was reported.

Author Contributions:
Conceptualization: Abdul-Karim; Data curation: Abukari Alhassan; Formal analysis: Abdul-Karim Iddrisu; Methodology: Abdul-Karim Iddrisu; Writing - original draft: Abdul-Karim Iddrisu; Writing - review & editing: Abukari Alhassan.

ABSTRACT

The National Research Council panel on prevention and treatment of missing data in clinical trials recommends that primary analysis methods are carefully selected before appropriate sensitivity analysis methods can be chosen. In this paper, we recommend an appropriate primary analysis method for handling CD4 count data from the IMPI trial and trials with similar settings. The estimand of interest in the IMPI trial is the effectiveness estimand hypothesis. We discussed, compared, and contrasted results from complete case analysis and simple imputation methods, with the direct-likelihood and multiple imputation methods. The simple imputation methods produced biased estimates of treatment effect. However, the maximum likelihood and the multiple imputation methods produced consistent estimates of treatment effect.  The maximum likelihood or the multiple imputation approaches produced unbiased and consistent estimates. Therefore, either the maximum likelihood or the multiple imputation methods, under the assumption that the data are missing at random can be considered as primary analysis method when one is considering sensitivity analysis to dropout using the CD4 count data from the IMPI trial and other trials with similar settings.

## 1. BACKGROUND

In longitudinal clinical trials, measurements are taken repeatedly at some selected occasions for each patient. Typically, a number of patients dropout for reasons such as adverse effect of treatment or protocol deviations [1]; which are often not under the control of the investigator. Since missing data may result in loss of vital information and reduction in the precision which is likely to biased inferences, it might be necessary to accommodate dropout in the modelling process in order to produce valid inference. To account for missing data in the modelling process, it is important to have an idea about the process generating the missingness process, known as missing data mechanisms [2], [3]. Missing data mechanisms can be classified into three main classes according to Rubin [4] taxonomy on reasons for missing data as missing completely at random (MCAR), missing at random (MAR), and not missing at random (NMAR).

Primary analysis for clinical trials

Data are missing completely at random (MCAR) if the probability of missingness is independent of the observed and unobserved outcomes of the variable being analysed. For instance, data missing due to administrative reasons could be classified as MCAR since the reason for missingness has nothing to do with the response model or the covariates. Data are missing at random (MAR) if the probability of missingness is independent of unobserved outcomes of the variable being analysed. Data missing due to lack of benefit (efficacy) from treatment could be attributed to MAR mechanism. This is because the reason for dropout is associated with the observed response and thus, in some sense, can be predicted from the response model [5]. Data are not missing at random (NMAR) if the probability of missingness is dependent on the observed and the unobserved outcomes of the variable being analysed. For instance, drop out after a sudden unobserved drop in treatment benefit (efficacy) could be classified as NMAR since the drop out will would be dependent on the unobserved and would not be predictable from the observed data alone.

Modelling of incomplete longitudinal data has been an active area of biostatistics research. It is known that the use of methods that do not account for the missing data may biased inferences [6]–[10] especially when data are not missing at random. One of such methods is the complete case analysis (CC). This method assumes that the data are missing at random and deletes all patients with missing values and then performs statistical analyses on the remaining patients with no missing data (complete cases).

There are three main and commonly used methods for handling missing data in longitudinal data analysis. These methods are the (1) imputation methods (2) likelihood-based methods, and (3) weighting methods. Generally speaking, these methods are interrelated in such a way that they ``impute'' certain values for missing data. The difference among these methods is that the imputation methods impute values for the missing values explicitly, whereas the likelihood-based and the weighting methods ``impute'' missing values implicitly. We evaluate the performance of the simple imputation approaches against the multiple imputation and the likelihood-based approaches using real data from the IMPI trial [11], [12]. Our aim is to recommend primary analysis method for analysis of the incomplete CD4 count data from the IMPI trial. This is necessary because it is based on the primary analysis method that proper sensitivity analysis can be formulated to asses sensitivity of the results under the primary analysis to alternative plausible assumptions about the missing data [13]–[15]. Sensitivity analysis to the incomplete CD4 count data from the IMPI can be found in [16]–[18].

We now describe the IMPI trial data and then point out why the trial aim is on effectiveness estimand. Thereafter, we discuss common issues associated with missing data and the three key recommendations by the National Research Council (NRC) [13] on prevention and treatment of missing data in clinical trials. These discussions are linked with the aim of the IMPI trial in terms of the estimand to address. We discuss some of the common methods for handling missing data. This discussion is necessary to identify an appropriate primary analysis method suitable for the CD4 count data. We then applied these methods to the CD4 count data and then discuss the results. We give concluding remarks where we recommend a primary analysis method for the CD4 count data from the IMP trial.

Primary analysis for clinical trials

## 1.1. Description of the IMPI trial data

In this paper, we used data from the IMPI trial [11], [12]. The IMPI trial is a multicentre international randomized doubled-blind placebo-controlled 2 x 2 factorial study. The IMPI trial tested prednisolone and Mycobacterium indicus pranii (M. indicus pranii) immunotherapy treatments in TB pericarditis patients in Africa. TB pericarditis leads to high mortality especially in countries with limited resource and with concomitant epidemics of human immunodeficiency virus (HIV) infection [11], [12]. Tuberculosis pericarditis is associated with high morbidity and mortality even if anti-tuberculosis treatment is taken as directed [12]. A reduction in the strength of the inflammatory response in TB pericarditis may improve patient's conditions by reducing cardiac tamponade and pericardial constriction. However, whether the use of adjunctive immunomodulation with corticosteroids and M. indicus pranii can safely reduce mortality and morbidity is uncertain [12]. To investigate whether adjunctive immunomodulation with corticosteroids and M. indicus pranii can safely reduce mortality and morbidity, Mayosi and colleagues set up the IMPI trial [11], [12].

In total, 1400 patients with definite probable tuberculosis pericardial effusion, from 9 African countries in 19 centers were enrolled in the four-year trial. Eligible patients were randomly assigned to receive oral pill prednisolone for 6 weeks and M. indicus pranii or placebo for 3 months. Patients were followed up at weeks 2, 4, and 6 and months 3 and 6 during the intervention period and 6-monthly thereafter for up to 4 years [11].

The main aim of the IMPI trial was to assess the effectiveness and safety of oral pill prednisolone and M.w injection in reducing the time to first occurrence of the primary composite outcome of death, pericardial constriction, or cardiac tamponade requiring pericardial drainage in with TB pericardial effusion [19]. In this paper, we assessed the effect of trial medications (prednisolone) on CD4 count changes over time. We focus on HIV+ patients only who have at least two CD4 count values observed. In the IMPI trial, patients who were confirmed HIV positive at the time of randomization or confirmed to be HIV positive during the trial were given a standard of care (ART) and their CD4 counts were measured at some visits. Mayosi and colleagues [19] results showed that the oral pill prednisolone and M. indicus pranii do not interact and hence, treatments arms were analyzed separately with their corresponding placebo arms. We considered analysis of the CD4 count measurements under the prednisolone treatment and its corresponding placebo arm only. However, one may decide to consider analysis under M. indicus pranii treatment and its corresponding placebo arm. The analysis of CD4 counts data is restricted to the mandated periods for CD4 counts measurements; baseline, week 2, months 1, 3 and 6. However, most South Africa centres continued to measure CD4 counts at months 24, 36 and 48 scheduled visit time. These data were excluded in this analysis. A majority of patients had unobserved CD4 counts with 72%, 84%, and 93% as missingness proportions for the months 24, 36 and 48, respectively.

### 1.1.1. Monotone data

The monotone CD4 count data consisted of 137 HIV positive patients. 64 were in the placebo arm and 73 in the prednisolone arm. The left panel of the Figure 1 shows profiles plots of the observed $\sqrt{(CD4)}$ count measurements for each HIV positive patient. Some of these patients dropped out at months 0.5, 1, and 3, after the baseline measurements are taken, whereas some patients completed the study with their values observed from baseline up to month 6. The right panel of the Figure 1 shows profiles plots of the mean $\sqrt{(CD4)}$ count measurements by treatment arms, where it can be

Primary analysis for clinical trials

observed that there is a slight reduction of CD4 count level for patients in the prednisolone arm compared with those in the placebo arm.

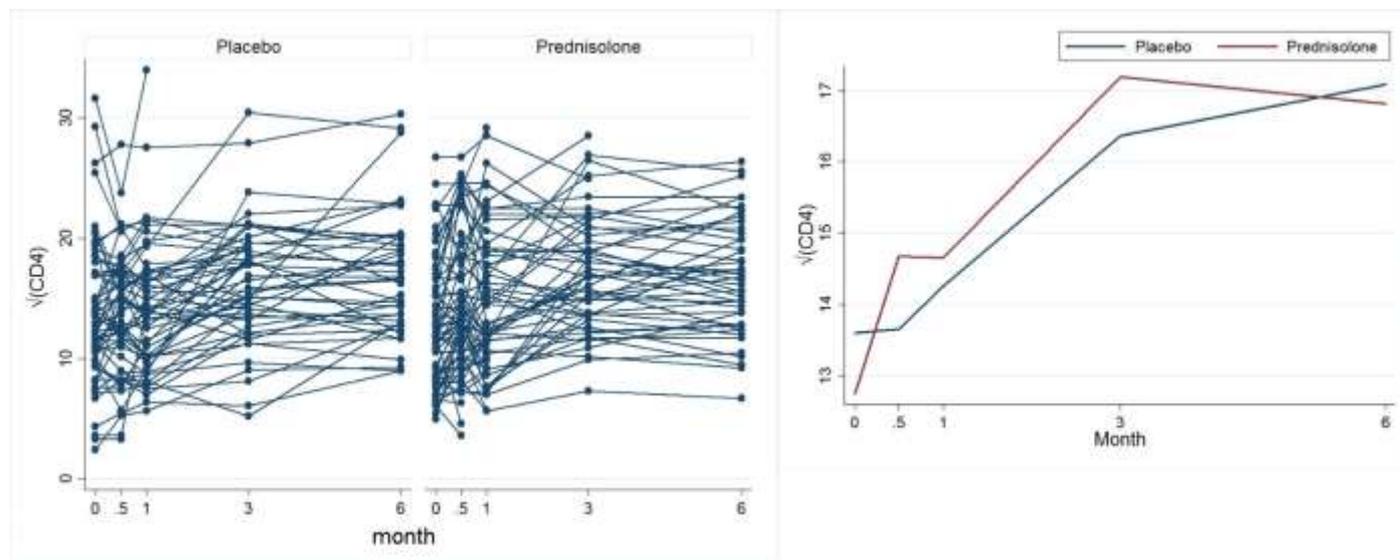

**Figure 1. Individual profiles plots of the monotone √(CD4) count data (left panel) and the mean √(CD4) count (right panel), by placebo and prednisolone treatment arms.**

Table 1 gives the number and proportion of patients remaining at each visit by treatment arms. There is higher completion rate 44 (69%) in the placebo arm compared with 46 (63%) completion rate in the prednisolone treatment arm. There are four deviations patterns. Table 2 shows the mean $\sqrt{(CD4)}$ count for each of the deviation pattern at each visit by treatment arm. The deviation patterns 4, 3, 2 and 1 represent completers (those patients who completed the study without missing values), those who dropped out at months 3, 1 and 0.5 respectively. The proportion (37%) of patients deviating is higher in the prednisolone arm compared with the proportion (31%) of patients deviating in the placebo. The distributions of the patterns of missingness between the two treatment groups do not differ (chi-squared test statistic = 5.15, p = 0.161).

### 1.2. Common issues associated with missing data and solutions

Randomized clinical trials are the recommended tool for evaluating the effect of new medical interventions [13]. However, it is often difficult to avoid occurrence of missing values in clinical trials and hence a substantial percentage of the measurements of the outcome of interest is often missing. This missingness reduces the benefit provided by the randomization and introduces potential biases in the comparison of the treatment groups [13].

Missing data can arise due to a variety of reasons, including the inability or unwillingness of participants to meet appointments for evaluation. Some of the reasons could be due to adverse effect of treatment or moving to partial compliance with treatment [15]. The National Research Council panel (NRC) concludes that a more principled approach to design and analysis in the presence of missing data is both needed and possible. In this study, we focused on the method of analysis of missing data.

We now focus on the three key recommendations by the NRC panel to deal with the missing data in the CD4 count data. These recommendations are (1) making precise and clear objectives of the trial, (2) minimizing the amount of missing data and (3) using plausible primary analysis together with sensitivity analyses [14], [15], [20]–[23]. Following these recommendations and in the context of the IMPI trial, this paper proposes an appropriate primary analysis method for the incomplete CD4 count data. Sensitivity analysis is beyond the scope of this paper; we only focused on selecting primary analysis method for the CD4 count data. Choosing a suitable primary analysis is crucial because a wrongly chosen primary analysis method will subsequently results in wrong sensitivity analysis formulations.

### 1.2.1. Objectives of the trial

The NRC [13] recommends the need to set out clear and precise objectives of the trial. This is a measure to avoid ambiguities in conclusions caused by missing data. The use of post deviation data largely depends on the estimand of interest. The debate on appropriate estimands is based on whether the focus is on efficacy or effectiveness [15], [24], [25][5], [24], [25]. For instance, if one is interested in the difference in outcome improvement at the planned end period for all randomized patients, missing post deviation data can be used in the primary analysis. In this scenario, the hypothesis to address is the effectiveness estimand hypothesis. This is because the effectiveness estimand compares treatment groups irrespective of what treatment patients received, and thus inferences is on the effectiveness of the treatment regimen and not the originally randomized treatment [14]. On the other hand, if one is interested in the difference in outcome improvement assuming that all patients adhered to treatment, then missing post deviation data cannot be used in the primary analysis. In this scenario, the efficacy estimand will be of interest since the estimand compares causal effects of the initially randomized treatment if taken as directed [14]. Furthermore, if the interest is in the difference in outcome improvement in all randomized patients at the planned end period of the trial attributable to the initially randomized treatment, missing post deviation or imputed data can be used. In this case the hypothesis to address is the effectiveness hypothesis. Whenever there is the need to use post deviation data, control imputation analyses are used as a means to obtained post deviation data needed to estimate effectiveness [15], [24], [25].

Following the above explanations on the objective the trial, it is now clear from the data description section that the hypothesis to address is the effectiveness hypothesis. This is because the trial aim is on outcome improvement at the planned end period for all randomized patients.

### 1.2.2. Minimizing the amount of missing data

The best approach to missing data is to avoid the occurrence of missing data. The development of new analytic methods and software tools for analyzing incomplete data has been an active area of research and more achievements have been made in that regards. However, all analyses are still challenged by the confusing and difficult problem of analyzing incomplete data. For instance, all analyses require assumptions about the missing data, these assumptions cannot be verified from the data, and the appropriateness of analyses and inference cannot be ensured. According to the NRC, minimizing missing data is the best means of handling the problem. In this paper, we address this need by using methods of analysis that predict or impute the missing values in the CD4 count data. In the context of the IMPI trial, the use of missing post deviation data, which we do not have, can be compensated by using methods that predict or estimate the missing data before data analysis.

### 1.2.3. Primary and sensitivity analyses

Another key recommendation is to use an appropriate primary analysis and sensitivity analyses approaches to assess the sensitivity of the primary analysis results to key assumptions about the missing data. In order to decide on appropriate primary analysis, the process generating the missing data must be taken into account in the statistical analysis.

The commonly used methods of analysis are the likelihood-based and multiple imputation (MI) [26], [27]. These methods can reduce the potential bias arising from missing data. The NRC panel encourages increased use of these methods and the assumptions underlying these methods need to be clearly communicated to medical experts so that they can assess their validity.

## 2. METHODS

This section introduces some of the approaches for drawing inferences in the presence of missing data. For a number of commonly used methods, users are not always aware of the assumptions that underlie the methods and the results drawn from applying them. This lack of awareness is particularly true of single imputation methods such as LOCF or BOCF and mixed effects regression models [28] that rely on strong parametric assumptions.

In principle, a decision has to be made concerning the modelling approach for the missing data process. There is no universal method for handling incomplete data in a clinical trial. Each trial has its own set of design and measurement characteristics. The trial designers should decide on a primary set of assumptions about the missing data mechanism. Those primary assumptions then serve as a reference for the sensitivity analyses. In many cases, the primary assumptions can be missing at random (MAR). Assumptions about the missing data mechanism must be transparent and accessible to clinicians. A statistically valid analysis should be conducted under the primary missing data assumptions.

We now discuss the methods for handling missing data and then point out the missingness mechanism required for each of these methods to yield valid inference. The advantages and disadvantages associated with using such methods are discussed.

### 2.1. Complete case (CC) analysis

The CC analysis is not a method for handling missing data because CC analysis confines analysis to only the completers [29] where completers are patients with no missing values. The CC analysis is valid when the mechanism of the missing data is assumed to be MCAR and statistical analyses can be performed using standard methods of analysis. Some of the disadvantages associated with using such analysis are small sample size, loss of statistical power due to the sample size, and biased inferences. The CC method may be preferred under a situation where the sample size is large, proportion of missing data is small, and missing data mechanism is MCAR [6]. Severe bias may occur when the missingness mechanism is MAR but not MCAR. This bias can be positive or negative, as illustrated by [30]. This approach may yield reliable estimates of treatment effect in a clinical trial setting where patients remain on treatment but investigators or clinicians could not measure their response to treatment at all scheduled visits time due to administrative reason.

### 2.2. Imputation-based approaches

Imputation of missing data is one of the approaches for handling missing data. The basic idea behind imputation approach is that missing values are filled with the imputed values to obtain complete data sets. One advantage of imputation approach is that, once filled-in data set has been obtained, standard methods for complete data can be applied. The two widely used imputation methods are the simple and multiple imputation methods.

#### 2.2.1. Simple imputation methods

Simple imputation (SI) methods explicitly impute a single value for the missing data and replace all unobserved responses with such value. The two widely used simple imputation method are the last observation carried forward (LOCF) and baseline observation carried forward (BOCF).

##### 2.2.1.1. Last observation carried forward method

The last observation carried forward method [29], [31] replaces missing values of dropouts in the study with their respective last observed values. This approach assumes that the response of a patient does not change after he or she drops out after randomization. This approach is questionable in many settings since dropouts might have lose some treatment benefit they have obtained while on treatment [29], [31]. For this approach to make sense, very strong and often unrealistic assumptions have to be made. This is because at times, one has to believe that a patient's measurements stay at the same level from the time of dropout onward or during the period they are unobserved.

The left panel of Figure 2 displays individual complete profile plots of $\sqrt{(CD4)}$ count data using the last observation carried forward method. As expected, the response profiles of patients who dropped out, of the study, have their profiles remained constant over the remaining period. The right panel of Figure 2 displays the mean $\sqrt{(CD4)}$ count data by placebo and prednisolone treatment arms. The mean profiles show an upward trend in the CD4 count evolution over time relative to the right panel of the Figure 1. The upward trend may be due to the constant CD4 count value after a patient dropped out. This trend suggests that the LOCF is likely to overestimate the unobserved CD4 count values since some patients' CD4 count values might have reduced (that is patients lost some benefit of treatment) after dropped out. As a result, the LOCF is likely to produce upward biased estimates of treatment effect. Verbeke and colleagues [32] showed that all components (fixed and random parts) of a linear mixed model may be severely affected by using the LOCF method. If one wishes to ``carry forward'' information after withdrawal, then the appropriate distribution should be carried forward, not the observation [23].

##### 2.2.1.2. Baseline observation carried forward

The BOCF approach replaces the missing values for each patient in the study with their respective baseline observations. This approach assumes that responses for dropouts remain the same as they were measured before randomization. This assumption is very intuitive in a clinical studies involving HIV+ patients or patients with chronic cancer [15]. However,

Primary analysis for clinical trials

this assumption might not always hold since most patients in a given study might have obtained significant benefit from the intervention after randomization.

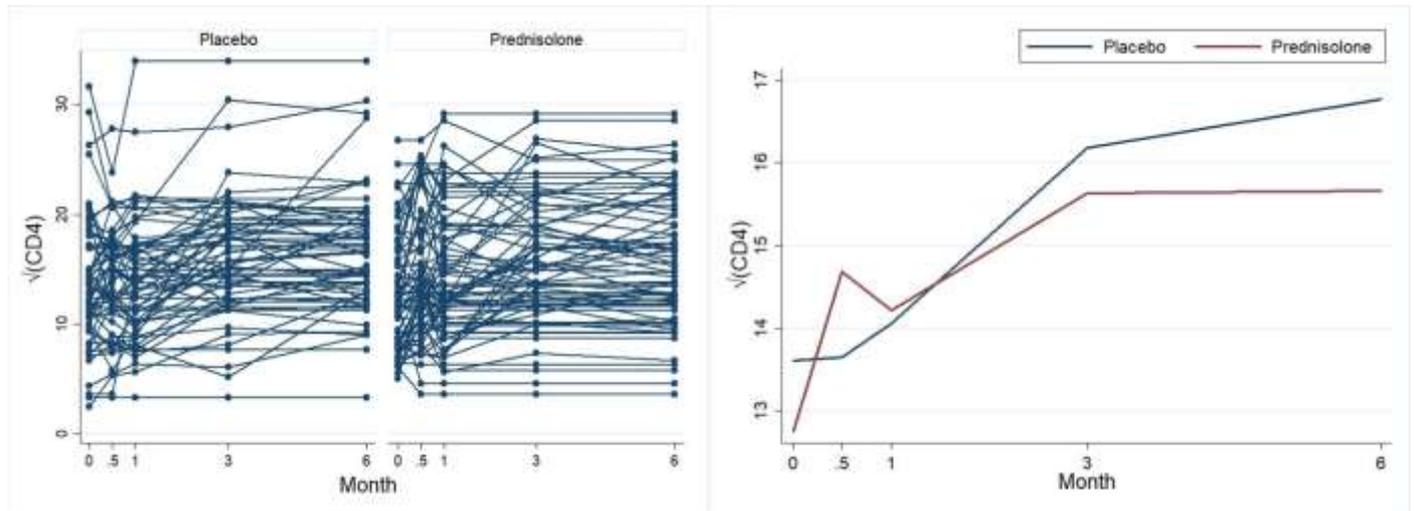

**Figure 2. LOCF: Individual profiles plots of the monotone √(CD4) count data (left panel) and the mean √(CD4) count (right panel), by placebo and prednisolone treatment arms.**

The left panel of Figure 3 displays individual complete profile plots of $\sqrt{(CD4)}$ count data using the baseline observation carried forward method. It can be observed that the response profiles of patients, who dropped out, of the study, have their profiles declined to the baseline value over the remaining period. This is expected because dropouts baseline values are used for the unobserved measurements in the remaining period. The right panel of Figure 3 displays the mean $\sqrt{(CD4)}$ count data by placebo and prednisolone treatment arms. The mean profiles show a downward trend in the CD4 count evolution over time relative to the right panel of the Figure 1 and Figure 2. This trend suggests that the BOCF is likely to underestimate the unobserved CD4 count values since some patients' CD4 count values would not necessarily reduce to the baseline value after drop out. As a result, the LOCF is likely to produce downward biased estimates of treatment effect.

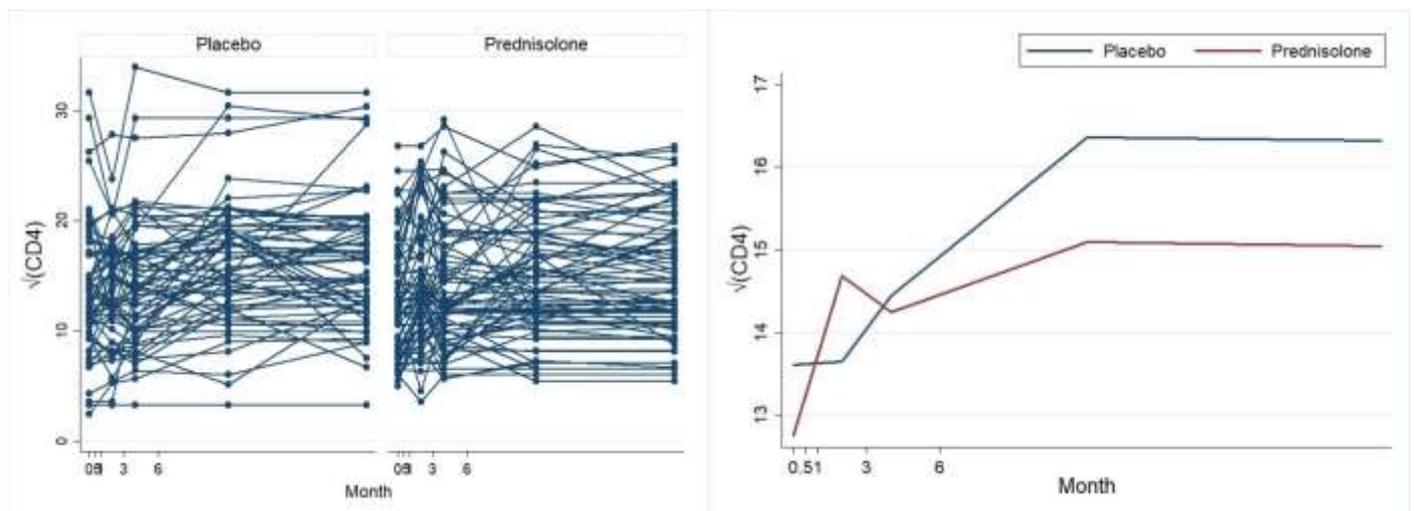

**Figure 3. BOCF: Individual profiles plots of the monotone √(CD4) count data (left panel) and the mean √(CD4) count (right panel), by placebo and prednisolone treatment arms.**

Primary analysis for clinical trials

Although, the LOCF and the BOCF methods may biased estimates of treatment effects and inflated rates of false positive and false-negative results, are likely [13], [14], [25], [31], the use of such methods set a historical example that, when combined with the desire to compare results with historical findings and the belief that LOCF and BOCF yielded conservative estimates of treatment effects, encouraged continued use of these methods [14]. An alternative approaches to the LOCF and BOCF limitations are the multiple imputation and the direct likelihood-based methods [29], [31].

### 2.2.2. Multiple imputation (MI)

Multiple imputation approaches assume that the data are missing at random (MAR) and create a set of K imputed values for the missing values there by creating K-complete data sets [4], [33]. MI accounts for the uncertainty inherent in the imputation of the unobserved responses.

Several imputation methods proposed drawing missing values $Y_i^m$ from the conditional distribution of unobserved responses $Y_i^m$, given the observed responses $Y_i^o$ and any potential covariate, $P(Y_i^m|Y_i^o, \boldsymbol{\theta})$, where $\boldsymbol{\theta}$ is a vector of parameter estimates describing the measurement process [26]. Let $\widehat{\boldsymbol{\theta}}$ be an estimator of the parameter $\boldsymbol{\theta}$ describing the complete data, with associated estimates $\widehat{\boldsymbol{\vartheta}}$ of the variance-covariance $\boldsymbol{\vartheta}$ matrices. This means that the K-completed data analyses corresponding to K-imputations under one model give rise to K complete-data statistics. These statistics are then combined to a single parameter estimates that appropriately adjust for non-responses under such imputation model. Thus, the values of $\widehat{\boldsymbol{\theta}}$ and $\widehat{\boldsymbol{\vartheta}}$ calculated on the K- completed data sets are $\theta_1, \cdots, \theta_k$, and $\vartheta_1, \cdots, \vartheta_k$ respectively.

Rubin [34] showed that multiple imputation relies on the Bayesian posterior distribution of the parameter $\boldsymbol{\theta}$, defined as $P(\boldsymbol{\theta}|Y_i^o) = \int P(\boldsymbol{\theta}|Y_i^o, Y_i^m) P(Y_i^m|Y_i^o) dY_i^m$, where $P(\boldsymbol{\theta}|Y_i^o, Y_i^m)$ is the average over the repeated imputation from $P(Y_i^m|Y_i^o)$. The MI approach is attractive because it can be highly efficient, even for small values of $K$ [4], [26], [35] and increasing the $k$ also improves the precision of the between-imputation covariance matrix.

### 2.3. Direct maximum likelihood analysis

Direct maximum likelihood (ML) methods can be applied to incomplete longitudinal data. These methods can also be applied to incomplete longitudinal data either after case deletion or after imputation of the missing observations. Under the case deletion, since missing values are no longer present in the completers or in the imputed data set, likelihood approaches are based on the full-data likelihood (1) of the complete data. On the contrary, for incomplete longitudinal data, any method within a likelihood framework would require working with the observed-data likelihood (2).

The traditional likelihood-based methods involve maximization of the full-data likelihood defined as

$$L^* \equiv L^*(\boldsymbol{\theta}, \boldsymbol{\psi} \mid \boldsymbol{Y}, \boldsymbol{R}) = \prod_{i=1}^{N} P(Y_i, R_i|\boldsymbol{\theta}, \boldsymbol{\psi}) = \prod_{i=i}^{N} P(Y_i^o, Y_i^m, R_i|\boldsymbol{\theta}, \boldsymbol{\psi}), \qquad (1)$$

where $\boldsymbol{\psi}$ is a vector of parameters describing the dropout process. However, in the presence of missing values, inference must be based on what is observed, and thus, the full-data likelihood $L^*$ (1) must be replaced by the observed-data

Primary analysis for clinical trials

likelihood $L$ (2), for which the individual likelihood contributions need to be integrated over the missing values defined as

$$L \equiv L(\boldsymbol{\theta}, \boldsymbol{\psi} \mid Y, R) = \prod_{i=1}^{N} \int P(Y_i^o, Y_i^m, R_i \mid \boldsymbol{\theta}, \boldsymbol{\psi}) \, dY_i^m \quad (2)$$

Under ignorability (where observed data and the missing data mechanism can be modelled separately), the observed-data likelihood simplifies into either (3) for MCAR or (4) for MAR [2], [29]. For MCAR, the missing data mechanism component is defined as $P(R_i \mid Y_i^o, Y_i^m, \boldsymbol{\theta}, \boldsymbol{\psi}) = P(R_i \mid \boldsymbol{\psi})$. Hence

$$P(Y_i^o, R_i \mid \boldsymbol{\theta}, \boldsymbol{\psi}) = P(Y_i^o \mid \boldsymbol{\theta}) P(R_i \mid \boldsymbol{\psi}) \quad (3)$$

Given model (3), we can now model the measurement and the missing data mechanisms separately to produce valid inferences since $R_i$ is independent of $Y_i^o$ and $Y_i^m$. For MAR, the missing data mechanism component is defined as $P(R_i \mid Y_i^o, Y_i^m, \boldsymbol{\theta}, \boldsymbol{\psi}) = P(R_i \mid Y_i^o, \boldsymbol{\psi})$ and hence

$$P(Y_i^o, D_i \mid \boldsymbol{\theta}, \boldsymbol{\psi}) = P(Y_i^o \mid \boldsymbol{\theta}) P(D_i \mid Y_i^o, \boldsymbol{\theta}, \boldsymbol{\psi}), \quad (4)$$

where $D_i$ is the occasion at which the ith patient dropped out. Given model (4), valid statistical inferences cannot be obtained without accounting for the missing data mechanism since $D_i$ depends on the observed $Y_i^o$.

Under MAR, the log-likelihood is defined as

$$\ell_i = \ln L_i = \ln P(R_i \mid Y_i^o, \boldsymbol{\psi}) + \ln P(Y_i^o \mid \boldsymbol{\theta}) \quad (5)$$

Under the MCAR, the log-likelihood is defined as

$$\ell_i = \ln L_i = \ln P(R_i \mid \boldsymbol{\psi}) + \ln P(Y_i^o \mid \boldsymbol{\theta}), \quad (6)$$

where the model for the non-response need not be conditioned on the observed responses. In this case, the direct maximum likelihood approach is no more complicated than fitting a likelihood-based model on the complete cases. Analysis under the NMAR mechanism is beyond the scope of this paper since our interest is to determine an appropriate primary analysis approach for handling the incomplete CD4 count data from the IMPI trial.

When missing data mechanism is described as MAR and ignorable, the likelihood-based methods are effectively imputing the missing values by modelling and estimating parameters for the joint distribution of the responses, $P(Y_i \mid X_i, \boldsymbol{\theta})$. Rubin [2] showed that likelihood-based inferences can be obtained by integrating over the missing responses from the joint distribution of the responses $P(Y_i \mid X_i, \boldsymbol{\theta})$, defined by

$$L(\boldsymbol{\theta}) \propto \prod_{i=1}^{N} \int P(Y_i^o, Y_i^m \mid X_i, \boldsymbol{\theta}) dY_i^m \quad (7)$$

Intuitively, the missing values $Y_i^m$ are validly predicted by the observed data via the model for conditional mean, $E(Y_i^m \mid Y_i^o, X_i, \boldsymbol{\theta})$ [36]. The expectation maximization (EM) algorithm [27] is used to obtain the ML estimate of $\boldsymbol{\theta}$ that maximizes the likelihood function (7).

3. **RESULTS**

Primary analysis for clinical trials

In this section, we applied the CC analysis, MI, ML, LOCF and the BOCF methods to the incomplete CD4 count measurements. To determine whether the dropout mechanism is likely to be MCAR, we carried out a logistic regression of the probability of completion on the time, prednisolone, and change in CD4 ($\Delta CD4$) after each visit and whether patient was on ART at each schedule visit (ART) variables. Table 3 shows the results obtained by fitting a mixed effect logistic model to the probability of dropout out. It can be observed that time, ART and $\Delta CD4$ are predictors of patient withdrawal in the IMPI trial. The change in response after each visit ($\Delta CD4$) shows that the missing data mechanism is unlikely to be MCAR since the MCAR assumption about missing data stipulates that the probability of dropout is independent of the observed and the unobserved.

The results from Little's MCAR test [37] also showed that the data are not missing completely at random. This means that the process that generated the missing data in the IMPI trial is likely to be MAR since probability of dropout depends of on the changes in observed values. We used the LOCF, BOCF and MI methods to obtain complete data sets. The linear mixed effect model (8) [38] was then fitted to the imputed data. The direct maximum likelihood (ML) method also assumed linear mixed effect model (LMM) for the observed data. Let $Y_i = (Y_{i1}, Y_{i2}, \cdots, Y_{in})$ denotes an N-dimensional vector of the responses for the ith patient and $X_i$ be an $N \times p$ design matrix of covariates for the ith patient. The linear mixed effect model (LMM) [38] is assumed for the measurement process and is given by

$$\begin{cases} Y_i = X_i\beta + Z_i b_i + \varepsilon_i, \\ b_i \sim N(0, G_i(\rho)), \\ \varepsilon_i \sim N(0, \Sigma_i(\sigma)), \\ \quad b_i \perp \varepsilon_N, \end{cases} \quad (8)$$

where $b_i$ is an $q$ dimensional vector of random effects, $Z_i$ and $X_i$ are $N \times q$ and $N \times q$ dimensional matrices of known covariates, $\beta$ is a $p$ vector containing the fixed effects, $\varepsilon_i$ is an $N$-dimensional vector of residual components, $G_i(\rho)$ and $\Sigma_i(\sigma)$ are $q \times q$ and $n \times n$ covariance matrices respectively and $\sigma$ and $\rho$ are $c \times 1$ and $s \times 1$ (with $s < n(n+1)/2$) vectors of unknown variance parameters corresponding to $\varepsilon_i$ and $b_i$ respectively.

Our fitted linear mixed model is defined as

$\sqrt{CD4}_{ij} = \beta_0 + \beta_i \times \text{prednisolone}_i + \beta_2 \times \text{month}_j + \beta_3 \times \text{prednisolone}_i \times \text{month}_j + \beta_4 \times \text{ART}_{ij} + \beta_5 \times \text{prednisolone}_i \times \text{ART}_{ij} + \beta_6 \times \text{Age}_i + b_i + \varepsilon_{ij}$, where $\sqrt{CD4}_{ij}$ the square root of CD4 counts is for ith patient at the jth visit, for $i = 1, \cdots, N$ and $j = 1, \cdots, n$, $b_{0i}$ represents the patient-specific random effect, and $\varepsilon_{ij}$ is the residual error. It is assumed that $b_i$ and $\varepsilon_{ij}$ are independently distributed as $b_i \sim N(0, \sigma_b^2)$ and $\varepsilon_{ij} \sim N(0, \sigma_\varepsilon^2)$ respectively. Also, the $\beta_1, \beta_2, \beta_3, \beta_4, \beta_5,$ and $\beta_6$ represent the effects of prednisolone (anti-tuberculosis treatment), time (months), prednisolone-time interaction, ART, prednisolone-ART interaction and Age respectively.

Table 4 shows the results of analyses from the CC, LOCF, BOCF, ML and MI methods using the CD4 count data from the IMPI trial. The results in Table 4 show that there is no significant prednisolone effect. The effect of prednisolone-ART interaction is also not significant. This confirms that prednisolone treatment does not influence the benefit of ART treatment. Patients CD4 count levels significantly increased with increasing time and patients who received ART at each visit have significantly higher CD4 count levels relative to those who did not receive ART at all schedule visit. The prednisolone-time interaction results show a very slight increase in CD4 count level in the placebo arm compared with prednisolone arm over time. However, this increase is not significant. The results also showed that older patients are more likely to have lower CD4 counts compared with the younger people.

Under the CC analysis, 90 (66%) of the 137 patients was used. The approach assumes that the missing data mechanism is MCAR. However, the results in Table 3 showed that data from the patients whose CD4 counts were unobserved in the IMPI trial is unlikely to be MCAR. This is because missingness is more likely to depend on the CD4 counts changes over time (MAR mechanism). This means that the unobserved data could substantially change the conclusions. For instance, in all analyses, CC analysis gives an upward biased estimate of treatment effect. Hence, using the CC analysis may not be sensible since it does not address any scientific question in the incomplete CD4 count measurements.

As expected, the LOCF method produced upward biased estimate of treatment effect compared with the BOCF method. This is because the LOCF assumes that CD4 counts measurements for dropouts remain constant over the remaining study period. However, this assumption is not plausible in the IMPI clinical trial setting. In the IMPI trial, we expect CD4 counts in either arm to decline or increase among some HIV+ patients. This implies that the LOCF method is likely to overestimate or inflate CD4 counts values for dropouts whose CD4 counts level declined at some scheduled visit times. This seems to explain why the LOCF produce upward biased estimates.

The BOCF produced downwards biased of treatment effect. This is expected because the BOCF assumes that the benefit of treatment after dropout is equivalent to benefit observed at the time of randomization. This assumption makes sense in general especially in clinical study designs on disease such as chronic cancer [15] or HIV+ patients. However, this is not the case in the IMPI trial. In the IMPI trial, patients did not drop out from treatment, rather, their CD4 count levels were not measured due to inadequate resources. This make the BOCF method also not suitable for handling the missing data in the CD4 count data from the IMPI trial. Hence, the BOCF methods cannot be used as the basis for primary analysis.

The ML analysis produced lower estimate of treatment effect compared with the MI analysis. The ML and MI analyses often provide similar estimates under assumptions that the missing data mechanism is MAR and the observed data is normally distributed. However, in a clinical trial setting where MAR is more likely and the MI imputation model uses the same variables as in the substantive model, the MI produces higher and efficient estimates.

4. **DISCUSSION**

In this paper, we proposed an appropriate primary analysis for analysing the CD4 counts data from the IMPI trial [12], [39] and other trials with similar settings. We pointed out that, under the IMPI trial, the hypothesis to address is the effectiveness hypothesis. This hypothesis estimates the effect of treatment if actually taken recognizing that patients may lose the benefit of treatment at the end period of the trial. We have discussed, compared, and contrasted some of the methods for handling missing data in clinical trials. These methods were illustrated using the CD4 counts data from the IMPI trial. Results from these methods showed that, there is no significant prednisolone effect, there is significant time effect, and there are no significant treatment-time and treatment-ART interactions effects. Patients who are on ART treatment are likely to have significantly high CD4 counts compared with those who are not on ART. The results also showed that older patients are more likely to have lower CD4 counts. In addition, the results showed that the data from the patients who withdrew from the IMPI trial is unlikely to be MCAR. Hence using the CC analysis is not suitable method. This is because the unobserved data could substantially change the conclusions.

Our analyses showed that both the ML and MI methods produced consistent estimates of the treatment effect. The MI approach to missing data addresses the limitations of the LOCF, CC, and the BOCF by accounting for the variability due to missing data. In conclusion, the MI under MAR mechanism seems to provide efficient estimates compared with the other missing data methods considered in this paper. We therefore recommend that either the MI or ML under MAR mechanism assumption can be considered as primary analysis method when one is considering sensitivity analysis [16]–[18] to dropout using the CD4 count data from the IMPI trial and other trials with similar settings.

We note that these analyses are based on the CD4 count data from the IMPI trial and hence methods identified in this paper, as recommended primary analysis methods, cannot be used as primary analysis methods in other trials. This is because trial objectives vary from trial to trial [13].

We also emphasis on the need to consider detail illustration of how primary analysis methods are chosen for a given trial when one's focus is on performing sensitivity of the primary analysis results to alternative plausible assumptions about the missing data. That is, even if there is evidence to show that the data collection from a given trial was random, inferences in the presence of missing data should illustrate to readers of the clinical report that the data was indeed missing at random. This will enable readers of the report to evaluate significance of the methods in relation to how they scientifically addressed the missing data from the trial under consideration.

It is important to carefully select a suitable primary analysis method, more especially, when focus is to investigate the sensitivity of the primary analysis results to alternative assumptions about the missing data mechanism. This is very crucial because a wrongly chosen primary analysis will subsequently results in wrong sensitivity analysis formulations. This will also lead to wrong statistical inferences.

Primary analysis for clinical trials

**Tables**:

Table 1. Percentage of patients remaining in the study at each visit

| Month | Placebo N (%) | Prednisolone N (%) |
|---|---|---|
| 0 | 64 (100) | 73 (100) |
| 0.5 | 64 (100) | 73 (100) |
| 1 | 57 (88) | 63 (86) |
| 3 | 53 (83) | 51 (70) |
| 6 | 44 (69) | 46 (63) |

Primary analysis for clinical trialsTable 2. $\sqrt{(CD4)}$ count at each visit by dropout pattern and treatment arm

| Dropout pattern* | Dropout time (month) | | | | | N (%) |
|---|---|---|---|---|---|---|
| | 0 | 0.5 | 1 | 3 | 6 | |
| Placebo | | | | | | |
| 4 | 13.14 | 13.47 | 13.62 | 16.24 | 17.09 | 44 (69) |
| 3 | 12.58 | 13.702 | 14.76 | 17.01 | - | 9 (14) |
| 2 | 16.90 | 17.68 | 20.27 | - | - | 4 (6) |
| 1 | 15.98 | 12.44 | - | - | - | 7 (11) |
| Mean (sd) | 13.61 (5.84) | 13.65 (4.97) | 14.26 (5.32) | 16.37 (4.85) | 17.09 (5.14) | 64 (100) |
| Prednisolone | | | | | | |
| 4 | 13.18 | 15.52 | 14.70 | 16.76 | 16.82 | 46 (63) |
| 3 | 19.89 | 19.27 | 19.51 | 21.17 | - | 5 (7) |
| 2 | 9.04 | 12.26 | 12.49 | - | - | 12 (16) |
| 1 | 11.64 | 11.43 | - | - | - | 10 (14) |
| Mean (sd) | 12.75 (5.10) | 14.68 (5.71) | 14.66 (5.84) | 17.19 (4.80) | 16.82 (4.72) | 73 (100) |

*Dropout patterns: 4 = subjects who had all measurements up to 6 months (completers), 3 = subjects who had measurements up to 3 months, 2 = subjects who had measurements up to 1 month, and 1 = subjects who had measurements up to 2 weeks.*

Primary analysis for clinical trials

Table 3. Adjusted odds ratios for withdrawal

| Variable | Odds Ratio | SE (95% CI) | P-value |
|---|---|---|---|
| Month | 0.85 | 0.013 (0.821, 0.873) | <0.001 |
| ART | 8.62 | 3.374 (3.999, 18.561) | <0.001 |
| $\Delta CD4$ | 1.24 | 0.026 (1.186, 1.239) | <0.001 |

Primary analysis for clinical trials

**Table 4. Parameter estimates from the methods of analyse**

| Method | Prednisolone | | | Time | | | Prednisolone x Time | | | ART | | | Prednisolone x ART | | | Age | | |
|---|---|---|---|---|---|---|---|---|---|---|---|---|---|---|---|---|---|---|
| | Est | Std | p-value | Est | std | p-value | Est | std | p-value | Est | std | p-value | Est | std | p-value | Est | std | p-value |
| CC | 1.31 | 0.962 | 0.1710 | 0.36 | 0.106 | <0.0001 | -0.12 | 0.14 | 0.4140 | 3.14 | 0.598 | <0.0001 | -0.43 | 0.834 | 0.6020 | -3.68 | 1.755 | 0.0360 |
| LOCF | 0.38 | 0.880 | 0.6640 | 0.32 | 0.082 | <0.0001 | -0.02 | 0.027 | 0.4320 | 2.83 | 0.485 | <0.0001 | -0.543 | 0.658 | 0.4090 | -3.01 | 1.605 | 0.0610 |
| BOCF | 0.08 | 0.876 | 0.9270 | 0.082 | 0.22 | 0.0090 | -0.04 | 0.027 | 0.1450 | 3.07 | 0.486 | <0.0001 | -0.08 | 0.660 | 0.9070 | -3.21 | 1.598 | 0.0450 |
| ML | 0.28 | 0.849 | 0.7430 | 0.39 | 0.102 | <0.0001 | -0.12 | 0.139 | 0.3850 | 2.98 | 0.543 | <0.0001 | -0.21 | 0.752 | 0.7760 | -3.14 | 1.548 | 0.0420 |
| MI | 0.56 | 0.791 | 0.4810 | 0.140 | 0.39 | 0.0050 | 0.13 | 0.195 | 0.4940 | 2.76 | 0.689 | <0.0001 | -1.12 | 0.966 | 0.2450 | -3.12 | 1.372 | 0.0230 |

**Acknowledgements**

We would like to thank the Mayosi Research Group, Department of Medicine, and University of Cape Town for providing the data for this paper.

[1] J. R. Carpenter, J. H. Roger, and M. G. Kenward, 'Analysis of Longitudinal Trials with Protocol Deviation: A Framework for Relevant, Accessible Assumptions, and Inference via Multiple Imputation', *J. Biopharm. Stat.*, vol. 23, no. 6, pp. 1352–1371, 2013.

[2] D. B. Rubin, 'Inference and missing data', *Biometrika*, 1976, doi: 10.1093/biomet/63.3.581.

[3] R. J. A. Little, 'Modeling the drop-out mechanism in repeated-measures studies', *J. Am. Stat. Assoc.*, 1995, doi: 10.1080/01621459.1995.10476615.

[4] D. B. Rubin, 'Inference and missing data', *Biometrika*, vol. 63, no. 3, pp. 581–592, 1976.

[5] B. Ratitch, M. O'Kelly, and R. Tosiello, 'Missing data in clinical trials: From clinical assumptions to statistical analysis using pattern mixture models', *Pharm. Stat.*, 2013, doi: 10.1002/pst.1549.

[6] J. on Kim and J. Curry, 'The treatment of missing data in multivariate analysis', *Sociol. Methods Res.*, 1977, doi: 10.1177/004912417700600206.

[7] R. J. A. Little, 'Regression with missing X's: A review', *J. Am. Stat. Assoc.*, 1992, doi: 10.1080/01621459.1992.10476282.

[8] S. Greenland and W. D. Finkle, 'A critical look at methods for handling missing covariates in epidemiologic regression analyses', *American Journal of Epidemiology*. 1995, doi: 10.1093/oxfordjournals.aje.a117592.

[9] J. L. Schafer and J. W. Graham, 'Missing data: Our view of the state of the art', *Psychol. Methods*, 2002, doi: 10.1037/1082-989X.7.2.147.

[10] X. Zhu, 'Comparison of Four Methods for Handing Missing Data in Longitudinal Data Analysis through a Simulation Study', *Open J. Stat.*, 2014, doi: 10.4236/ojs.2014.411088.

[11] B. M. Mayosi et al., 'Rationale and design of the investigation of the management of pericarditis (IMPI) trial: A 2 × 2 factorial randomized double-blind multicenter trial of adjunctive prednisolone and Mycobacterium w immunotherapy in tuberculous pericarditis', *Am. Heart J.*, 2013, doi: 10.1016/j.ahj.2012.08.006.

[12] B. M. Mayosi et al., ' Prednisolone and Mycobacterium indicus pranii in Tuberculous Pericarditis ', *N. Engl. J. Med.*, 2014, doi: 10.1056/nejmoa1407380.

[13] N. R. Council, *The Prevention and Treatment of Missing Data in Clinical Trials*. 2010.

[14] C. Mallinckrodt et al., 'Recent Developments in the Prevention and Treatment of Missing Data', *Ther. Innov. Regul. Sci.*, 2014, doi: 10.1177/2168479013501310.

[15] J. R. Carpenter, J. H. Roger, and M. G. Kenward, 'Analysis of longitudinal trials with protocol deviation: A framework for relevant, accessible assumptions, and inference via multiple imputation', *J. Biopharm. Stat.*, 2013, doi: 10.1080/10543406.2013.834911.

[16] I. Abdul-Karim and F. Gumedze, 'An application of a pattern-mixture model with multiple imputation for the analysis of longitudinal trials with protocol deviations', *BMC Med. Res. Methodol.*, 2019, doi: 10.1186/s12874-018-0639-y.

[17] I. Abdul-Karim and F. Gumedze, 'Application of sensitivity analysis to incomplete longitudinal CD4 count data', *J. Appl. Stat.*, pp. 1–16, 2018.

[18] I. Abdul-Karim and F. Gumedze, 'Sensitivity analysis for the generalized shared-parameter model framework', *J. Biopharm. Stat.*, pp. 1–19, 2019.

[19] B. M. Mayosi et al., 'Prednisolone and Mycobacterium indicus pranii in tuberculous pericarditis', *N. Engl. J. Med.*, vol. 371, no. 12, pp. 1121–1130, 2014.

[20] G. Molenberghs, G. Verbeke, H. Thijs, E. Lesaffre, and M. G. Kenward, 'Influence analysis to assess sensitivity of the dropout process', *Comput. Stat. Data Anal.*, 2001, doi: 10.1016/S0167-9473(00)00065-7.

[21] M. G. Kenward, 'Selection models for repeated measurements with non-random dropout: An

illustration of sensitivity', *Stat. Med.*, 1998, doi: 10.1002/(SICI)1097-0258(19981215)17:23<2723::AID-SIM38>3.0.CO;2-5.

[22] G. Verbeke, G. Molenberghs, H. Thijs, E. Lesaffre, and M. G. Kenward, 'Sensitivity analysis for nonrandom dropout: A local influence approach', *Biometrics*, 2001, doi: 10.1111/j.0006-341X.2001.00007.x.

[23] J. R. Carpenter, M. G. Kenward, and I. R. White, 'Sensitivity analysis after multiple imputation under missing at random: A weighting approach', *Stat. Methods Med. Res.*, 2007, doi: 10.1177/0962280206075303.

[24] C. H. Mallinckrodt, Q. Lin, and M. Molenberghs, 'A structured framework for assessing sensitivity to missing data assumptions in longitudinal clinical trials', *Pharm. Stat.*, 2013, doi: 10.1002/pst.1547.

[25] B. T. Ayele, I. Lipkovich, G. Molenberghs, and C. H. Mallinckrodt, 'A multiple-imputation-based approach to sensitivity analyses and effectiveness assessments in longitudinal clinical trials', *J. Biopharm. Stat.*, 2014, doi: 10.1080/10543406.2013.859148.

[26] D. B. Rubin, 'Multiple Imputation after 18+ Years', *J. Am. Stat. Assoc.*, 1996, doi: 10.1080/01621459.1996.10476908.

[27] A. P. Dempster, N. M. Laird, and D. B. Rubin, ' Maximum Likelihood from Incomplete Data Via the EM Algorithm ', *J. R. Stat. Soc. Ser. B*, 1977, doi: 10.1111/j.2517-6161.1977.tb01600.x.

[28] N. M. Laird and J. H. Ware, 'Random-effects models for longitudinal data', *Biometrics*, vol. 38, no. 4, pp. 963–974, 1982.

[29] R. J. A. Little and D. B. Rubin, *Statistical Analysis with Missing Data*. 2002.

[30] G. Molenberghs *et al.*, 'Analyzing incomplete longitudinal clinical trial data', *Biostatistics*, vol. 5, no. 3, pp. 445–464, 2004.

[31] G. Molenberghs and M. G. Kenward, *Missing Data in Clinical Studies*. 2007.

[32] G. Verbeke, Geert and Molenberghs, *Linear mixed models in practice: a SAS-oriented approach*. Springer Science & Business Media, 2012.

[33] D. B. Rubin, 'Multiple imputation after 18+ years', *J. Am. Stat. Assoc.*, vol. 91, no. 434, pp. 473–489, 1996.

[34] D. B. Rubin, 'The calculation of posterior distributions by data augmentation: Comment: A noniterative sampling/importance resampling alternative to the data augmentation algorithm for creating a few imputations when fractions of missing information are modest: The SIR', *J. Am. Stat. Assoc.*, vol. 82, no. 398, pp. 543–546, 1987.

[35] D. B. Rubin, 'Formalizing subjective notions about the effect of nonrespondents in sample surveys', *J. Am. Stat. Assoc.*, vol. 72, no. 359, pp. 538–543, 1977.

[36] G. Fitzmaurice, M. Davidian, G. Verbeke, and G. Molenberghs, *Longitudinal data analysis*. CRC Press, 2008.

[37] C. Li, 'Little's test of missing completely at random', *Stata J.*, 2013, doi: 10.1177/1536867x1301300407.

[38] N. M. Laird and J. H. Ware, 'Random-Effects Models for Longitudinal Data', *Biometrics*, 1982, doi: 10.2307/2529876.

[39] B. M. Mayosi *et al.*, 'Rationale and design of the Investigation of the Management of Pericarditis (IMPI) trial: A 2$\times$ 2 factorial randomized double-blind multicenter trial of adjunctive prednisolone and Mycobacterium w immunotherapy in tuberculous pericarditis', *Am. Heart J.*, vol. 165, pp. 109–115, 2012.